\documentclass[fleqn]{annalen}
\usepackage{graphics}
\usepackage{amssymb}
\begin{document}
\newcommand{\volume}{11}              
\newcommand{\xyear}{2000}            
\newcommand{\issue}{5}               
\newcommand{\recdate}{15 November 1999}  
\newcommand{\revdate}{dd.mm.yyyy}    
\newcommand{\revnum}{0}              
\newcommand{\accdate}{dd.mm.yyyy}    
\newcommand{\coeditor}{ue}           
\newcommand{\firstpage}{507}         
\newcommand{\lastpage}{510}          
\setcounter{page}{\firstpage}        
\newcommand{\keywords}{cosmology: gravitational lensing, dark matter, large
scale structure of universe} 
\newcommand{\PACS}{95.35.+d, 98.62.Sb, 98.65.Cw}
\newcommand{\shorttitle}
{R. Kaufmann et al., Giant Arc Statistics and Cosmological Parameters} 
\title{Giant Arc Statistics and Cosmological Parameters}
\author{R.\ Kaufmann$^{1}$ and N.\ Straumann$^{1}$} 
\newcommand{\address}
  {$^{1}$Institut f\"ur Theoretische Physik der Universit\"at Z\"urich-Irchel, 
   Z\"urich, Switzerland}
\newcommand{\email}{\tt mercator@physik.unizh.ch, norbert@physik.unizh.ch} 
\maketitle
\begin{abstract}
We study with semi-analytical methods the statistics of pronounced arcs caused
by lensing of galaxies by foreground galaxy clusters. For the number density and redshift
distribution of rich clusters we use Press-Schechter theory, normalized 
on the basis of empirical data. For the background sources we make use of 
observational results in the Hubble Deep Field. We present results for three
different lens models, in particular for the universal profile suggested by
Navarro, Frenk and White. Our primary concern is the dependence of the
expected statistics on the cosmological parameters, $\Omega_M$, $\Omega_\Lambda$.
The theoretical estimates are compared with the cluster arcs survey EMSS,
and the resulting constraints in the $\Omega$-plane are presented. In
spite of considerable theoretical an observational uncertainties a low-density
universe is favored. Degeneracy curves for the optical depth and
likelihood regions for the arc statistics in the $\Omega$-plane depend only
weakly on the cosmological constant.
\end{abstract}

\section{\label{introduction}Introduction}
There is now significant evidence that we are living in a critical universe
in which the vacuum energy (or some effective equivalent) dominates and 
ordinary (baryonic) matter is only a tiny fraction of the total matter content:
\begin{displaymath}
\Omega_\Lambda\simeq\frac{2}{3},\qquad \Omega_M\simeq \frac{1}{3}, \qquad
\Omega_M\gg\Omega_B.
\end{displaymath}
The first two statements are strongly favored by the combination of recent
measurements of the luminosity-redshift relation for type Ia supernovae at 
high redshifts and the observed temperature fluctuations of the cosmic
microwave background radiation. (For a review, see \cite{bahcall}.)
$\Omega_M\simeq 0.3\pm0.1$ is also consistent with data from rich clusters of
galaxies, in particular from the analysis of the X-radiation emitted by the
hot intra-cluster gas. The total matter contribution of rich clusters can also 
be determined by making use of (strong and weak) gravitational lensing.
Such observations provide an estimate of the ratio $\Omega_B/\Omega_M$.
$\Omega_B$ and $\Omega_M$ can then be obtained separately
by using in addition the primordial abundances of light elements, synthesized
during the first three minutes. (For a discussion of other methods, we refer
again to \cite{bahcall} and references therein.)

Readers who do not closely follow these developments should, however, be warned
that there remain serious worries. For instance, one
cannot yet exclude systematic effects in the supernovae data (e.g., intrinsic evolution,
extinction) which could masquerade a cosmological constant.

In this situation any method which provides independent restrictions
on the density parameters $\Omega_M$, $\Omega_\Lambda$ is of 
great interest. Lensing statistics on large scales has this potential.
The use of gravitational lensing as a tool to determine cosmological parameters
has been  suggested  in the pioneering works \cite{refsdal, press}. An early 
detailed study of lensing statistics was performed by Turner, Ostriker and Gott
\cite{turner}. Much of the extensive later work is based on 
this paper. After some sporadic studies which investigated the effect
of the cosmological constant on some specific aspects of lensing,
it was pointed out in \cite{fukugita, turner2, fukugita2} that
statistical properties of gravitational lensing may exhibit strong 
$\Lambda$-dependencies.

This is not the place to review the many more recent studies in the field
of lensing statistics. We confine ourselves to the following remarks (more
extensive references can be found in the quoted papers):

(i) Several authors have recently re-analyzed the statistics
of strong gravitational lensing of distant quasars by galaxies
\cite{kochanek, chiba, chiba2, cheng}. Observationally, there are only a
few strongly lensed quasars among hundreds of objects. The 
resulting bounds on $\Omega_M$ and $\Omega_\Lambda$ are, however, not very tight
because of systematic uncertainties in the galaxy luminosity functions, dark
matter velocity dispersions, galaxy core radii and limitations of the observational material.

(ii) On the basis of existing surveys, the statistics of strongly lensed radio 
sources has been studied in several recent papers \cite{falco, cooray, helbig}.
Beside some advantages for constraining the cosmological model,
there is the problem that the redshift distribution of the radio
sources is largely unknown. (One can, however, make use of a strong
correlation between the redshift and flux density distributions.)

(iii) Clusters with redshifts in the interval 
$0.2\lesssim z_c\lesssim 0.4$ are efficient lenses for
background sources at $z_s\sim 1$. For several reasons one can expect that the
probability for the formation of pronounced arcs is a sensitive function
of $\Omega_M$ and $\Omega_\Lambda$. First, it is well-known 
that clusters form earlier in low density universes. Secondly,
the proper volume per unit redshift is larger for low density universes and depends
strongly on $\Omega_\Lambda$ for large redshifts. An extensive numerical study
of arc statistics has been performed by Bartelmann et al.\ \cite{bartelmann},
with the result that the optical depth depends strongly on $\Lambda$.
In our semi-analytical treatment, discussed below, we find, however, only a
weak $\Lambda$-dependence.
 
This paper is organized as follows. Since most relativists are not
familiar with lensing statistics, we repeat
in section \ref{generalities} some basic concepts (cross sections,
optical depth) and formulae needed in the theoretical analysis. For a
comparison of theoretical expectations with observations, one has to
include magnification effects and the observational
selection criteria have to be taken into account. This will be explained
in section \ref{bias}. In section \ref{input} we discuss further inputs
we shall need in our semi-analytical treatment. This includes the
Press-Schechter formula for the differential comoving cluster number
density, dynamical and structural information required for the lens models,
as well as redshift and luminosity distributions for the background sources.
We shall compare our theoretical expectations with data for the Extended
Medium-Sensitivity Survey (EMSS), which will briefly be presented in section
\ref{emss}. Our main results are described in section \ref{results}.
These are compatible with a similar independent investigation
of Cooray \cite{cooray2}, but do not agree with the claimed 
$\Lambda$-dependence of Bartelmann et al.\ \cite{bartelmann}. 
Possible sources of this conflict will be indicated.
In section \ref{conclusions} we summarize and discuss future prospects
for tighter constraints in the $\Omega$-plane.

\section{\label{generalities}Generalities on lensing statistics}
As for any scattering situation, the concept of cross-sections
is basic in statistical lensing.

Consider  a source at redshift $z_s$ and distance $D_s$, as well as 
a lens a distance $D_d$ (we always use angular diameter distances). We are interested
in lensing events characterized by a set of lensing properties, $Q$, of
this gravitational lens system. In this paper we shall consider especially lensing
events in which a galaxy behind a cluster lens is seen as an arc with a length-to-width
ratio $R$, larger than some given value $R_0$. For a given lens at distance
$D_d$, the \emph{cross section} $\hat\sigma_{\scriptscriptstyle Q}$  for the properties $Q$
is by definition the area of the source sphere of radius $D_s$, within which a source
has to be located in order to be imaged with property $Q$.

It is convenient to also introduce the corresponding cross section $\sigma_{\scriptscriptstyle Q}$ in the
lens plane by a simple rescaling: $\sigma_{\scriptscriptstyle Q}=(D_d/D_s)^2\hat\sigma_{\scriptscriptstyle Q}$.

Consider next a distribution of lenses (deflectors) with number density $n_d(z_d)$  
at redshift $z_d$. The \emph{optical depth} $\tau_{\scriptscriptstyle Q}$ for property $Q$ is 
defined to be the probability that a given source at redshift $z_s$ to 
undergo a lensing event with property $Q$. This is equal to the fraction of the 
entire source sphere of radius $D_s$ which is covered by the cross sections
$\hat\sigma_{\scriptscriptstyle Q}$ (the cross-sections are in practice not overlapping):
\begin{equation}
\label{eq:depth}
\tau_{\scriptscriptstyle Q}(z_s)=\frac{1}{4\pi D_s^2}\int_0^{z_s}n_d(z_d)\left|\frac{dV(z_d)}{dz_d}\right|
\hat\sigma_{\scriptscriptstyle Q}(z_d)\,dz_d.
\end{equation}
Here, $dV/dz$ denotes the proper volume per unit redshift.
For a Friedmann universe we have, in standard notation,
\begin{equation}
\label{eq:dVdz}
\frac{dV}{dz}=4\pi D_d^2\left|\frac{dt}{dz}\right|,
\end{equation}
with
\begin{equation}
\left|\frac{dt}{dz}\right|=\frac{1}{H_0}\frac{1}{(1+z)E(z)},
\end{equation}
where
\begin{equation}
E^2(z)=\Omega_M(1+z)^3+(1-\Omega_M-\Omega_\Lambda)(1+z)^2+\Omega_\Lambda.
\end{equation}

In connection with the geometrical factor (\ref{eq:dVdz}) in the integral
(\ref{eq:depth}) we recall that the angular diameter distance $D(z)$
for a Friedmann universe is given by
\begin{equation}
D(z)=\frac{1}{H_0(1+z)}\frac{1}{|\Omega_K|^{1/2}}\mathcal{S}\left(|\Omega_K|^{1/2}
\int_0^z\frac{dz'}{E(z')}\right),
\end{equation}
where $\Omega_K:=1-\Omega_M-\Omega_\Lambda$ is a curvature parameter and 
$\mathcal{S}(\chi)=\sin\chi$, $\chi$, $\sinh\chi$ for $K>0$, $0$, $<0$, 
respectively. The dependence of $dV/dz$ on the cosmological
parameters is one of several sources for such a dependence of the optical
depth. Others are implicit in the cross-sections and the lens distributions.
Astrophysical uncertainties of these quantities limit
attempts to exact cosmological parameters from lensing statistics.

Inserting (\ref{eq:dVdz}) into (\ref{eq:depth}), the optical depth
can be expressed more simply in terms of the cross-section relative to the
lens sphere:
\begin{equation}
\label{eq:opticaldepth}
\tau_{\scriptscriptstyle Q}(z_s)=\int_0^{z_s}n_d(z_d)\sigma_{\scriptscriptstyle Q}(z_d)|dt/dz_d|\,dz_d.
\end{equation} 
Simple applications of this formula can be found in the
early papers on lensing statistics. (See also \cite{schneider},
especially chapters 11 and 12.)

We now consider in more detail the statistics of arcs produced by
cluster lenses. Let $n(M,z)\,dM$ be the comoving number density  of clusters
at redshift $z$ in the mass interval $dM$ about $M$. The cross-section
for the property that a background galaxy at redshift $z_d$ behind a cluster
of mass $M$ at redshift $z$ is seen as an arc with length-to-width ratio
larger than $R_0$ will be denoted by $\sigma(R_0,M,z,z_s)$.
In terms of these quantities the optical depth (\ref{eq:opticaldepth})
is given by
\begin{equation}
\label{eq:clusterdepth}
\tau(\geq R_0,z_s)=\int_0^{z_s}dz\,|dt/dz|(1+z)^3\int dM\, n(M,z)
\sigma(R_0,M,z,z_s).
\end{equation}

Singular isothermal spheres (SIS) are the simplest lens models we shall use. For these
the ratio $R$ coincides with the magnification $\mu$, and the cross-section is
\cite{schneider}
\begin{equation}
\label{eq:sis}
\sigma(\mu)=16\pi^3\left(\frac{\sigma_v}{c}\right)^4\left(\frac{D_dD_{ds}}{D_s}\right)^2
\frac{1}{(\mu-1)^2}.
\end{equation}
Here $\sigma_v$ is the (constant) one-dimensional velocity dispersion, and $D_{ds}$
the angular diameter distance from the lens to the source.

\section{\label{bias}Amplification bias and selection}
For any magnitude limited sample of sources, the number of
lensed sources is larger than it would be in an unbiased 
sample, because lensing brightens into the sample sources that 
would otherwise not be detected. This is, for example, a 
particularly pronounced effect in quasar lensing surveys, because 
the faint end of the quasar luminosity function rises steeply.

Taking the amplification bias into account, the lensing
probability (\ref{eq:clusterdepth}) has to be replaced by
\begin{equation}
\label{eq:prob}
\mathrm{prob}(\geq R_0,z_s,m_s)=\int_0^{z_s}\frac{d\tau}{dz_d}
B(m_s,z_s,z_d)\,dz_d,
\end{equation}
where the first factor under the integral is the integrand in
(\ref{eq:clusterdepth}) and the bias factor $B$, as a function of the apparent
source magnitude $m_s$ and the redshifts $z_s$, $z_d$, can be expressed
as follows: Let $dN_s/dm_s$ denote the unlensed number
counts of sources, which agrees in practice with the \emph{observed}
number counts. Then
\begin{equation}
\label{eq:bias}
B(m_s,z_s,z_d)=(dN_s/dm_s)^{-1}\int_{\Delta_{\mathrm{min}}}^\infty 
\frac{dN_s}{dm_s}(m_s+\Delta,z_s)q(\Delta)\,d\Delta,
\end{equation}
where $q(\Delta)\,d\Delta$ is the probability distribution for the
magnification of the apparent magnitude of lensed images.
Since $\Delta=2.5\log_{10}\mu$, the latter is simply related to the
magnification probability distribution $p(\mu)\,d\mu$. For a SIS model
we have
\begin{equation}
p(\mu)=\frac{2}{(\mu-1)^3}, \qquad \mu_{\mathrm{min}}=2.
\end{equation}
$dN_s/dm_s$ is simply related to the luminosity function of the
source population, for which a Schechter form is often adopted.

In optical arc searches only galaxies brighter than some limiting
apparent magnitude $m_{\mathrm{lim}}$ are counted. For a comparison
of theory with observation we must include this observational selection
in (\ref{eq:prob}) and (\ref{eq:bias}). Details of how this was implemented
are described in \cite{kaufmann}.

\section{\label{input}Additional theoretical and observational input}
For the numerical evaluation of (\ref{eq:prob}) and the corresponding
giant arc statistics we need explicit knowledge of cross-sections,
the differential comoving number density $n(M,z)$ of clusters,
and the distribution $(dN_s/dm_s)(m_s,z_s)$ of the source galaxies.
These quantities are, unfortunately, not well-known.

We have used three simple lens models, namely isothermal
spheres with vanishing and positive core radii, and also
the universal density profile proposed by Navarro, Frenk and White \cite{navarro}.
For isothermal models we need the velocity dispersion $\sigma_v$ that 
enters  in the cross-section with the fourth power (see (\ref{eq:sis})).
$\sigma_v$ depends on the cluster mass as well as on the cluster redshift $z$.
For the $z$-dependence  we adopt the one of the spherical
collapse model, while the $M$-dependence is derived from empirical
correlations, using in particular the correlation between 
$\sigma_v$ and the cluster temperature \cite{wu}.
(As expected from the virial theorem, the temperature
is strongly correlated with the mass, $T\propto M^{2/3}$.)

For $n(M,z)$ we use the Press-Schechter formula \cite{pressschechter}.
This is based on a mixture of statistical reasoning, linear perturbation
theory, and the spherical collapse model. Numerical simulations have 
demonstrated that the Press-Schechter expression provides
a much better description than could be expected. One of 
the assumptions that goes into its derivation is that the smoothed
density field is Gaussian. The function $n(M,z)$ is essentially
determined by the rms density fluctuations $\sigma(M,z)$ of \emph{linear}
theory. (For textbook treatments, see \cite{paddi, peacock}.)
The $z$-dependence of $\sigma(M,z)$ is then equal to the one of the 
growing mode of linear perturbation theory,
\begin{equation}
\mathcal{D}_g(z;\Omega_M,\Omega_\Lambda)=\frac{5}{2}
\Omega_M E(z)\int_z^\infty\frac{1+z'}{E^3(z')}\,dz',
\end{equation}
and $\sigma(M,z=0)$ can be expressed in terms of the power
spectrum of the density fluctuation in linear theory. For the latter
we use the CDM transfer function given by Bardeen et al.\ \cite{bardeen},
and assume that the initial fluctuations are adiabatic
and nearly scale free. The normalization parameter $\sigma_8$ is
taken from \cite{pen}. Uncertainties of this parameter, as well
as its dependence on $\Omega_M$, $\Omega_\Lambda$, are critical,
since $\sigma_8$ enters exponentially in the Press-Schechter formula.

For the source distribution of galaxies we make use of the
present knowledge based on the Hubble Deep Field (HDF). The 
deep exposure of  this survey is important for our analysis.
Unfortunately, the redshift distribution is not precisely
known, since a detailed spectroscopy of about 900 faint
sources in the sample is impossible. There exist, however,
two photometric redshift catalogs that were obtained by
analyzing the observations in four broadband filters. We 
are always using the catalog in \cite{sawicki} which is
based on spectral template fitting.

As a result of large K-corrections, the luminosity function is not
well-known, and cannot be used for determining the amplification bias.
Our method of treating this is discussed in \cite{kaufmann}.

The previous discussion should make it amply clear that we have to live 
with substantial uncertainties. Hopefully, these will be reduced in the 
near future.

\section{\label{emss}Extended Medium-Sensitivity Survey and arc statistics}
In the next section we shall compare our theoretical expectations
with the results of a CCD imaging survey of gravitational lensing
selected from the Einstein Observatory Extended Medium-Sensitivity Survey
(EMSS). The latter is a large and sensitive X-ray catalog and has been studied by
a number of groups, including followed up observations using the ROSAT or ASCA
satellites. Presently, about 100 EMSS sources are classified as clusters.

In \cite{luppino} a subsample was selected for an arc survey,
using the University of Hawaii 2.2 m telescope, and also the CFHT for 
specific observations. It should be emphasized that the EMSS is not
strictly flux limited, since the
detection of sources was optimized for high surface brightness.
Therefore, more extended clusters with lower surface brightness,
especially relatively close ones, may have been missed. Moreover,
at high redshifts only the most luminous clusters
can be seen. Since these are likely to be more massive, the lensing
fraction is expected to become larger.
For these reasons clusters in the redshift range $0.15<z<0.6$, and surface
brightness $L_x\geq 4\times10^{44}$ erg/s in the energy band
$0.3-3.5$ keV are most suitable for a comparison with theoretical expectations.

The subsample selected in \cite{luppino} contains 21 such clusters. (This
corresponds to about 1300 such clusters over the entire sky.) Among these
there are 6 clusters with giant arcs satisfying the following
criteria: (i) The length-to-width ratio $R$ of the arc is $\geq 10$. (ii) The
magnitude in the V-band satisfies $m_V\leq 22$. (Actually, $m_V$ was not
always measured; in those cases $m_R\leq 22$ was adopted, which
should be about the same.) The 6 clusters in the limited area of the sample
scale to about 360 over the entire sky, and the chance that a cluster in the
specified range of $z$ and $L_x$ contains an arc is roughly 0.3.

So far the statistics are poor, but with the new satellites CHANDRA and XMM the
situation should soon improve.

\section{\label{results}Results}
In this section we present the main results of our study. 
A more detailed discussion and additional parameter studies
are given in \cite{kaufmann}.
\begin{figure}[t!]
\centerline{\resizebox{10cm}{8.25cm}{\includegraphics{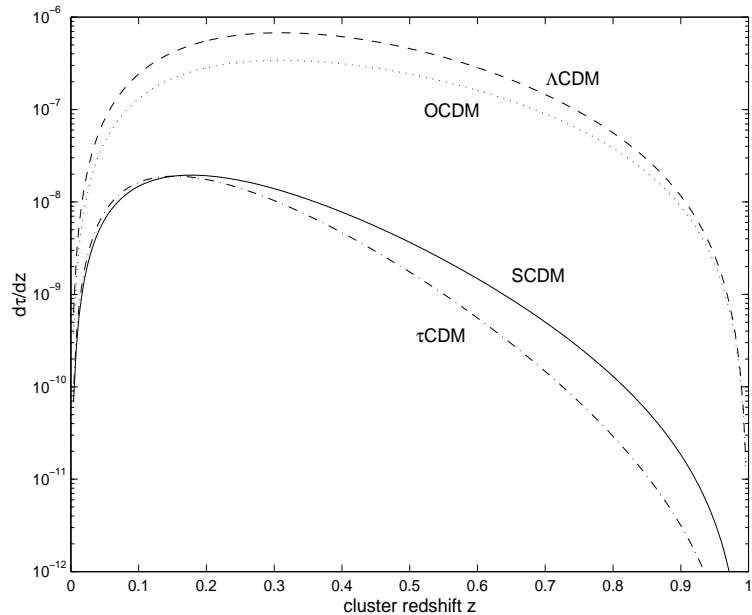}}}
\caption{\label{fig:depth}Differential optical depth for $z_s=1$ and $R\geq 10$.
Clusters are modeled as SIS. See text for details.}
\end{figure}

Fig.\ \ref{fig:depth} shows the differential optical depths for the
formation of giant arcs with $R\geq 10$ for the following four
cosmological models: SCDM ($\Omega_M=1$, $\Omega_\Lambda=0$),
OCDM ($\Omega_M=0.3$, $\Omega_\Lambda=0$), $\Lambda$CDM
($\Omega_M=0.3$, $\Omega_\Lambda=0.7$), $\tau$CDM ($\Omega_M=1$, $\Omega_\Lambda=0$,
shape parameter $\Gamma=0.23$). (The latter model mimics a scenario with an 
unstable massive $\tau$-neutrino.) The source  is taken at redshift
$z_s=1$. The foreground cluster lenses with X-ray luminosities in the
EMSS band $L_x\geq2\times 10^{44}$ erg/s, are modeled as singular
isothermal spheres.

Note - and this is one of our main points - that the curves for 
OCDM and $\Lambda$CDM do not differ much. The total 
optical depths are listed in the left column of
Table \ref{tab:results}, while the right column contains
those of the numerical study of Bartelmann et al.\ \cite{bartelmann}.
We emphasize that the agreement is quite good for three of the
four models, but for the OCDM model there is a difference of
about an order of magnitude. Possible sources of this discrepancy 
will be discussed later.
\begin{table}[b!]
\caption{Optical depth. Comparison between our results \cite{kaufmann} and the
results in \cite{bartelmann}.} 
{\small
\begin{center}
\begin{tabular}{lcc}                               
\\ \hline \\[-3mm]
\hspace*{-1.5mm}&$\tau$ \cite{kaufmann} & $\tau$ \cite{bartelmann}
\\ \hline \\[-3mm]
\hspace*{-1.5mm}SCDM&$3.1\times 10^{-8}$&$4.4\times 10^{-8}$ \\ 
\hspace*{-1.5mm}OCDM&$2.9\times 10^{-7}$&$2.9\times 10^{-6}$ \\
\hspace*{-1.5mm}$\Lambda$CDM&$5.3\times 10^{-7}$&$3.3\times 10^{-7}$ \\ 
\hspace*{-1.5mm}$\tau$CDM&$1.6\times 10^{-8}$&$4.4\times 10^{-8}$ \\  
\hline \\[-3mm] 
\end{tabular}
\end{center}
} \label{tab:results}
\end{table}

\begin{figure}[t!]
\centerline{\resizebox{10cm}{9.5cm}{\includegraphics{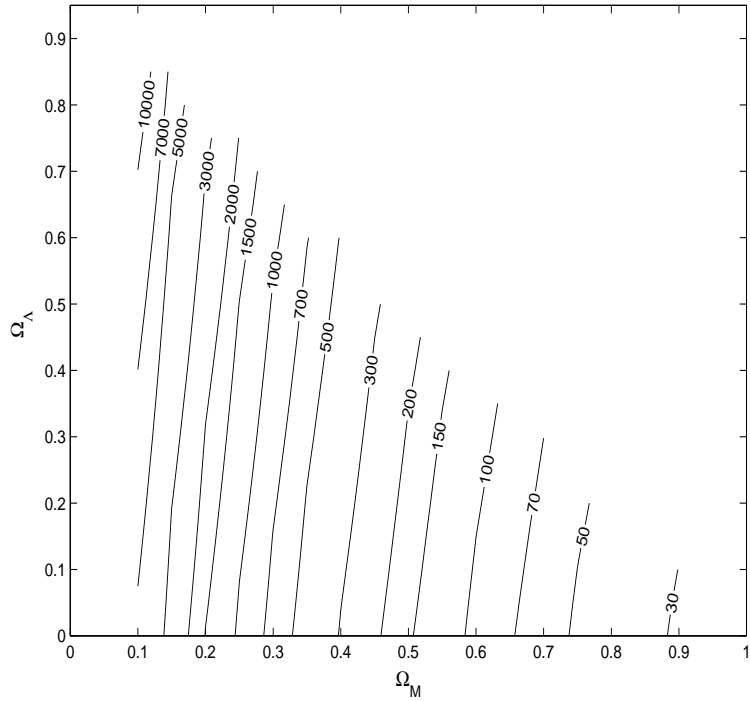}}}
\caption{\label{fig:number}Expected total number of giant arcs. 
$L_x\geq 4\times 10^{44}$ erg s${}^{-1}$, $m_V\leq 22$, $h_0=0.7$.}
\end{figure}

The dependence of the expected total number of giant arcs on the
cosmological parameters $\Omega_M$ and $\Omega_\Lambda$ 
is shown in Fig.\ \ref{fig:number}. Parameter variations and 
changes of the lens model have convinced us that the shape 
of the degeneracy curves (the lines of constant arc number)
is quite stable, and we believe that their weak $\Omega_\Lambda$-dependence
will also be found  in future more accurate studies.
The absolute numbers refer to the EMSS subsample specified a the end
of the previous section, and should be
compared with the 360 arcs mentioned there. The best agreement is reached
for $\Omega_M\approx 0.4$, practically independent of $\Omega_\Lambda$.

\begin{figure}[t!]
\centerline{\resizebox{10cm}{9cm}{\includegraphics{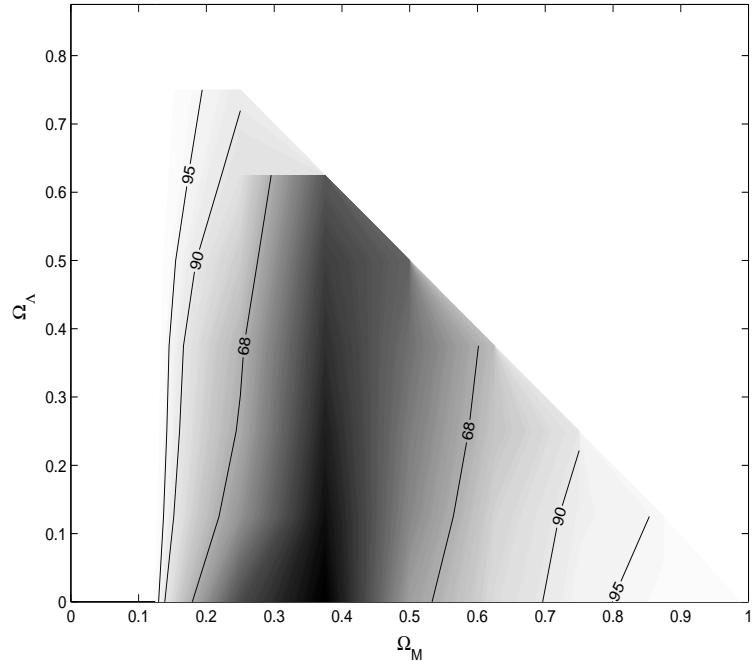}}}
\caption{\label{fig:likelihood}Likelihood contours.}
\end{figure}

In Fig.\ \ref{fig:likelihood} we present the result of a maximum
likelihood analysis, always keeping a SIS model. We have
marginalized with respect to several empirical
parameters, including $\sigma_8$, which affect the prediction most
strongly. At the present stage the 90\% likelihood regions are
still quite broad, but a low-density universe is clearly favored.

We have not yet made a similar likelihood analysis for the universal
profile of Navarro, Frenk and White \cite{navarro}. As already emphasized,
the degeneracy curves are quite similar (the slopes become slightly negative),
however, the absolute numbers turn out to be smaller \cite{kaufmann}. Other
profiles have recently been proposed; their implications for the arc-statistics will
be a subject of further studies.

As already mentioned, Bartelmann et al.\ \cite{bartelmann} found in their
N-body simulations for the optical depth a
strong $\Omega_\Lambda$-dependence (see Table \ref{tab:results}).
It is, of course, true that a numerical approach provides in principle
more realistic lensing properties of the clusters, since substructure
and asymmetries are included. What we find, however, difficult to understand
is the claim in \cite{bartelmann} that clusters are less concentrated
for the $\Lambda$CDM model than for the OCDM model with the same $\Omega_M$,
and that this is the main reason for the lower number of arcs in the former case.
A more recent simulation of another subgroup \cite{thomas} of the Virgo
consortium indeed found no substantial differences of the
cluster structures and mass profiles.

We have investigated the various $\Lambda$-dependences
which combine to the pattern of degeneracy lines shown
in Fig.\ \ref{fig:number}. The geometrical factor in (\ref{eq:clusterdepth})
is, of course, common to all lensing statistics, and the
$\Lambda$-dependence implicit in the SIS cross-section (\ref{eq:sis})
is quite typical. For a conserved comoving lens number density, these
would combine to degeneracy curves with $\Omega_\Lambda\approx\Omega_M$.
(This kind of degeneracies are, for instance, found for the statistics
of strongly lensed quasars.) This is changed by the $\Lambda$-dependence
of the Press-Schechter formula for $n(M,z)$ to the weak dependence in
Fig.\ \ref{fig:number}. (For further details, see \cite{kaufmann}.)

\section{\label{conclusions}Conclusions and outlook}
Statistical strong lensing data for a specific class of sources
have the potential to constrain the cosmological  parameters
$\Omega_M$ and $\Omega_\Lambda$. In combination with other 
information (from type Ia SNe, CMB anisotropies, etc.) this
could tighten the allowed range of these important 
cosmological parameters, and also serve as a consistency check.

So far, observational and theoretical uncertainties are limiting the use of 
lensing statistics for cosmology. Beside much better statistical samples we
also need accurate knowledge about the population and mass
distributions of the lenses, as well as on the redshift and luminosity
distributions of the sources.

In our semi-analytical study of giant arc statistics
we have investigated the range of the current systematic
uncertainties in constraining the cosmological parameters
in the $\Omega_M$-$\Omega_\Lambda$ plane. The available data for
EMSS favor a low-density universe, but the 90\% upper bound
for $\Omega_M$ is not really tight. More secure, we believe, is our
conclusion that the degeneracy curves in the $\Omega$-plane
depend only weakly on $\Lambda$. In contrast to other claims
in the literature \cite{bartelmann}, statistical lensing of giant arcs
thus can not be used to constrain $\Omega_\Lambda$. It may, however,
serve as a consistency check. (Note that the steep degeneracy
curves are semi-transversal to those for type Ia SNe, as well as to
the ones related to the position of the first acoustic peak in the
CMB anisotropies.)

One may be sceptical that strong lensing statistics
will ever become an accurate tool. Weak lensing by
large-scale structures looks more promising. The latter
produce a tiny gravitational shear field which
will be detectable in the near future. This shear pattern
is correlated with the power spectrum of density fluctuations.
The theoretical analysis is in this case more reliable than
for strong lensing statistics, because linear perturbation
theory is quite accurate on large scales. Weak lensing 
maps may soon be obtained with wide field
imaging surveys. (For a review, and references, see \cite{mellier}.)

\vspace*{0.25cm} \baselineskip=10pt{\small \noindent N.\ S.\  thanks the organizers
of the \emph{Journ\'ees Relativistes 1999} for inviting him to beautiful Weimar
and to present this work. He especially thanks G.\ Neugebauer for his gracious
hospitality. N.\ S.\ thanks M.\ Bartelmann for a useful discussion on
arc statistics. We are grateful to C.\ Frenk for providing us his FORTRAN subroutine
for the calculation of the universal profile parameters.

\noindent This work was supported by the Swiss National Science Foundation.}

\end{document}